\documentclass[reprint,amssymb,amsmath,aip,cha]{revtex4-1}
\usepackage{graphicx}
\usepackage[caption = false]{subfig}
\usepackage{color}
\usepackage{epsfig}
\usepackage{ifpdf}
\usepackage{url}%
\usepackage{cleveref}
\usepackage{bm}
\usepackage[colorlinks=true,linkcolor=blue]{hyperref}%
\expandafter\ifx\csname package@font\endcsname\relax\else
 \expandafter\expandafter
 \expandafter\usepackage
 \expandafter\expandafter
 \expandafter{\csname package@font\endcsname}% 
\fi
\begin{document}
\title{Experimental observation of self excited co--rotating multiple vortices in a dusty plasma with inhomogeneous plasma background.}
\affiliation{Institute for Plasma Research, HBNI, Bhat, Gandhinagar, 382428, 
India}
\author{Mangilal Choudhary}
\email{mangilal@ipr.res.in}
\author{S. Mukherjee}
\author{P. Bandyopadhyay}
%\affiliation{Institute for Plasma Research, Bhat, Gandhinagar, 382428, India}
%\author{}
% \date{5-12-2016}
%
\begin{abstract}
We report an experimental observation of multiple co--rotating vortices in a extended dust column in the background of non--uniform  diffused plasma. Inductively coupled RF discharge is initiated in the background of argon gas in the source region which later found to diffuse in the main experimental chamber. A secondary DC glow discharge plasma is produced to introduce the dust particles into the plasma. These micron sized poly-disperse dust particles get charged in the plasma environment and transported by the ambipolar electric field of the diffused plasma and found to confine in the potential well, where the resultant electric field of the diffused plasma (ambipolar E--field) and glass wall charging (sheath E--field) hold the micron sized particles against the gravity. Multiple co--rotating (anti--clockwise) dust vortices are observed in the dust cloud for a particular discharge condition. The transition from multiple to single dust vortex is observed when input RF power is lowered. Occurrence of these vortices are explained on the basis of the charge gradient of dust particles which is orthogonal to the ion drag force. The charge gradient is a consequence of the plasma inhomogeneity along the dust cloud length. The detailed nature and the reason for multiple vortices are still under investigation through further experiments, however, preliminary qualitative understanding is discussed based on characteristic scale length of dust vortex. There is a characteristic size of the vortex in the dusty plasma so that multiple vortices is possible to form in the extended dusty plasma with inhomogeneous plasma background. The experimental results on the vortex motion of particles are compared with a theoretical model and found some agreement.
\end{abstract}
\maketitle
\section{Introduction}
Dusty plasma is a low-temperature plasma consists of electrons, ions, neutrals, and sub--micron to micron sized
particles of solid matter (dielectric or conducting). When these dust particles are introduced into the conventional plasma, they undergo collisions with the highly mobile electrons more frequently than with the slower  and heavier ions within the plasma. As a result, the dust particles collect the negative charges upto $10^3-10^5$ times of an electronic charge. The interaction of theses highly negatively charged particles leads to exhibit the collective behavior because of the coulomb interaction. The instabilities \citep{instability1,instability2,instability3} in the medium provides the energy to grow the local/infinitesimal perturbation in this dissipative medium. Evolution of local perturbation appeares in the form of linear and non-linear dust acoustic modes \citep{daw2,daw3,dasw,pdasw,exp1dasw,mangilalpop}, dust lattice wave \citep{dlw1,dlw2}, and dust vortices \citep{vortexmicrogravity,largescalevortices,bellanicedustyrotation}. The vortex structures in the dusty plasma, which is one of the examples of the dynamical structures, are results of the collective response of the medium. These dynamic structures are mainly established either by dust particles motion or driven motion of plasma species (electrons and ions) in the dissipative medium. Properties of the dynamic structures, driven by the ions or electrons, changes while the external electric field or magnetic field is applied. The vortex or rotational motion of dust particles is widely studied in various dusty plasma systems. The spontaneous rotation of dust grains \citep{Agarwalrotation}, two--dimensional (2D) dust vortex flow \cite{uchida2dflow}, cluster rotation \cite{clusterrotationunmagnetizedplasma} poloidal rotation of dust grains with toroidal symmetry\citep{manjeetrotation}, and wave motion along with vortex motion \cite{selfexcitedmotioninhomogeneus,selfexcitedmotion,inductivelycoupledrotation} are observed in unmagnetized plasmas world wide. The horizontal and vertical vortex motion of dust grains in the presence of an auxiliary electrode near the levitated dust cloud in capacitive coupled plasma is also reported by Samarian \textit{et al.}\citep{horizontalrotation} and Law \textit{et al.}\citep{probeinducedcirculation}. The rotation or vortex motion of the dust particulates in the absence of magnetic field can be induced by asymmetric ions flow or sheared flow along with electric field \citep{rotationinionflow,vortexmicrogravity,laishramshearflow}, charge gradient of the particles along with the non--electrostatic forces \citep{vaulinajetp,vaulinaselfoscillation,selfexcitedmotion,zhdanovnonhamiltonian}, Rayleigh--Taylor instability \citep{rtinstabilityvortices} and transient shear instability \citep{transientinstabilityrotation}. \par
In the presence of magnetic field ($<$ 500 G), flows of electrons and ions influence the state of dusty medium and affect the dust cloud motion to rotate as a rigid body \citep{vasilmagneticrotation,
 angularvelocitysaturation,dustroataionmagnetickarasev,dynamicsinmagnetizedplasmasato,
 dzlievarotationstratamagnetic,magnetirotationecrplasma,knopkamagneticrotation}. Application of strong magnetic field ($\geqslant 4$ Tesla) can magnetized the micron sized dust particles, which may gyrate in a plane perpendicular to the magnetic filed vector \cite{thomasmpedx,thmasmagnetizeddustyplasma}. Apart from the magnetic field, neutral flow under some specific conditions can induce the dust mass rotation.  The dust rotation under the action of convective motion of background neutral gas, is studied by Ivlev et al. \citep{thermalconvection}. In the case of gas convection, dust rotation is setup because of the neutral--drag or thermophoretic force to dust grains \citep{thermalcreeprotation}. All the previous reported work suggests that dust dynamics is strongly affected by the motion of background species (electrons, ions and neutral) and can give different equilibrium dynamical structures.\par
Dynamics of the dusty plasma medium in inhomogeneous plasma environment is still unexplored area of research. Inhomogeneity in the plasma density and the electron temperature can triggers the various instabilities \cite{instability1,icpddw,vaulinaselfoscillation} in the dusty plasma medium, which excites the various dust acoustic modes and dust vortex motion. The dynamic structures (vortices) in the dusty plasma with inhomogeneous plasma background has been the subject matter of the present studies.\\
 In the paper, we report the onset of multiple co--rotating dust vortices in a dusty plasma medium. Dust particles are found to transport and trap in a potential well created by inductively coupled diffused plasma. Particles are confined in a combined electric field of diffused plasma (ambipolar E--field) and wall charging (sheath E--field). Various self--oscillatory motions  of dust particles such as acoustic vibrations and vortex motion are observed at different discharge parameters. In a parametric regime, when acoustic vibrations are diminished, multiple co--rotating (in anti--clockwise direction) vortices are found to be appeared in the dusty plasma medium. Occurrence of a stable dynamical structure (vortex) in dusty plasma exists due to potential sources which compensates the dissipative energy losses. The possible energy source to drive the vortex motion is charge gradient with electric field and the ion drag force. The dust vortex has a characteristics size in the dusty plasma with inhomogeneous plasma environment which depends on various dusty plasma parameters such as dust--dust interactions, dust density etc.. The experimental results shows the appearance of  multiple vortices when the dimension of the dust cloud becomes multiple of this vortex size.The qualitative understanding is discussed in the light of a theoretical model \cite{vaulinajetp,selfexcitedmotion,vaulinaselfoscillation} based on characteristic scale length of dust vortex in the dusty plasma and it appears that multiple vortices is possible to form in the extended dusty plasma with inhomogeneous plasma background. The experimental results on the vortex motion of particles are found in some agreement with the model. However, the detailed nature and the reason for multiple vortices are still under investigation through further experiments, as expanding Helical trajectories of dusts may not be ruled out, because the camera having limited depth of focus due to limited illuminated zone, sees a 2D image of a 3D structure. 
\par
The manuscript is organized as follows: Section~\ref{sec:exp_setup} deals with the detailed description on the experimental setup, plasma and dusty plasma production and their characterization. The experimental observations of co--rotating multiple dust vortices and their dynamics are discussed in Section~\ref{sec:plasma}. Quantitative analysis of the origin of multiple dust vortices in the dusty plasma medium is described in Section~\ref{sec:dust vortice}. A brief summary of the work along with concluding remarks is provided in Section~\ref{sec:conclusion}.
\section{Experimental Setup and Diagnostics} \label{sec:exp_setup}
\begin{figure*}%[h]
 \centering
\subfloat{{\includegraphics[scale=0.750]{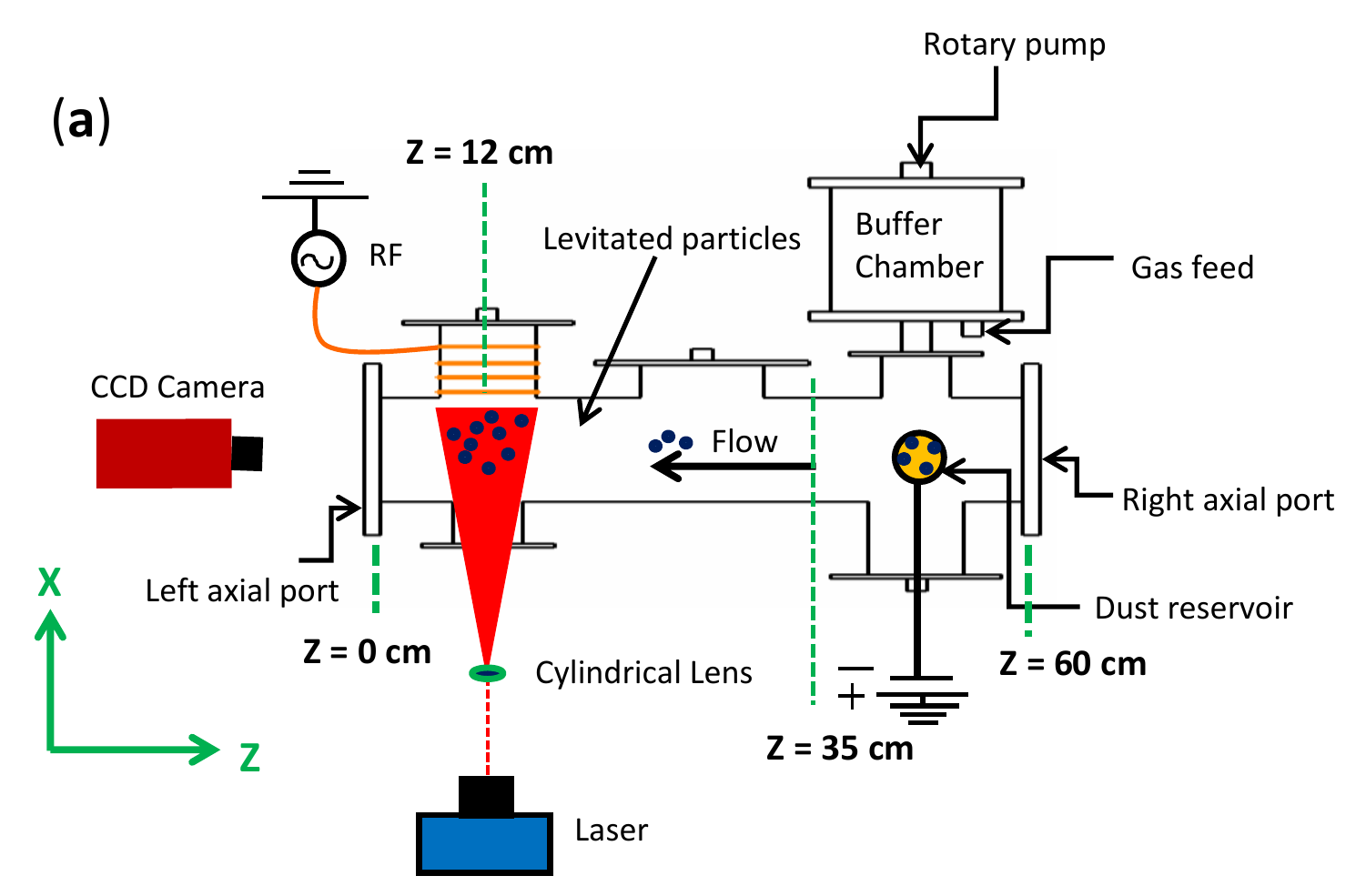}}}%
 \qquad
 \subfloat{{\includegraphics[scale=0.300]{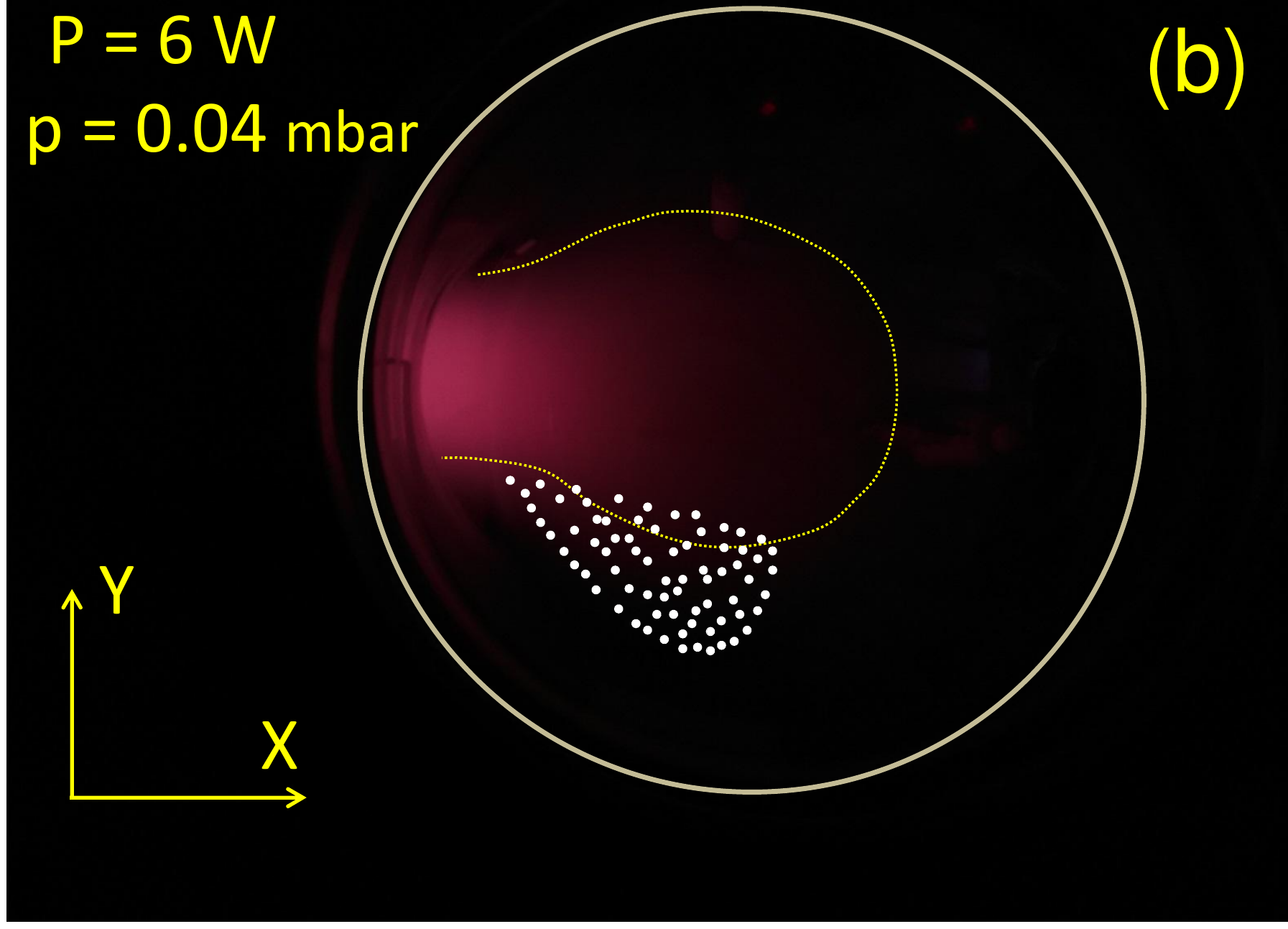}}}%
 \subfloat{{\includegraphics[scale=0.325]{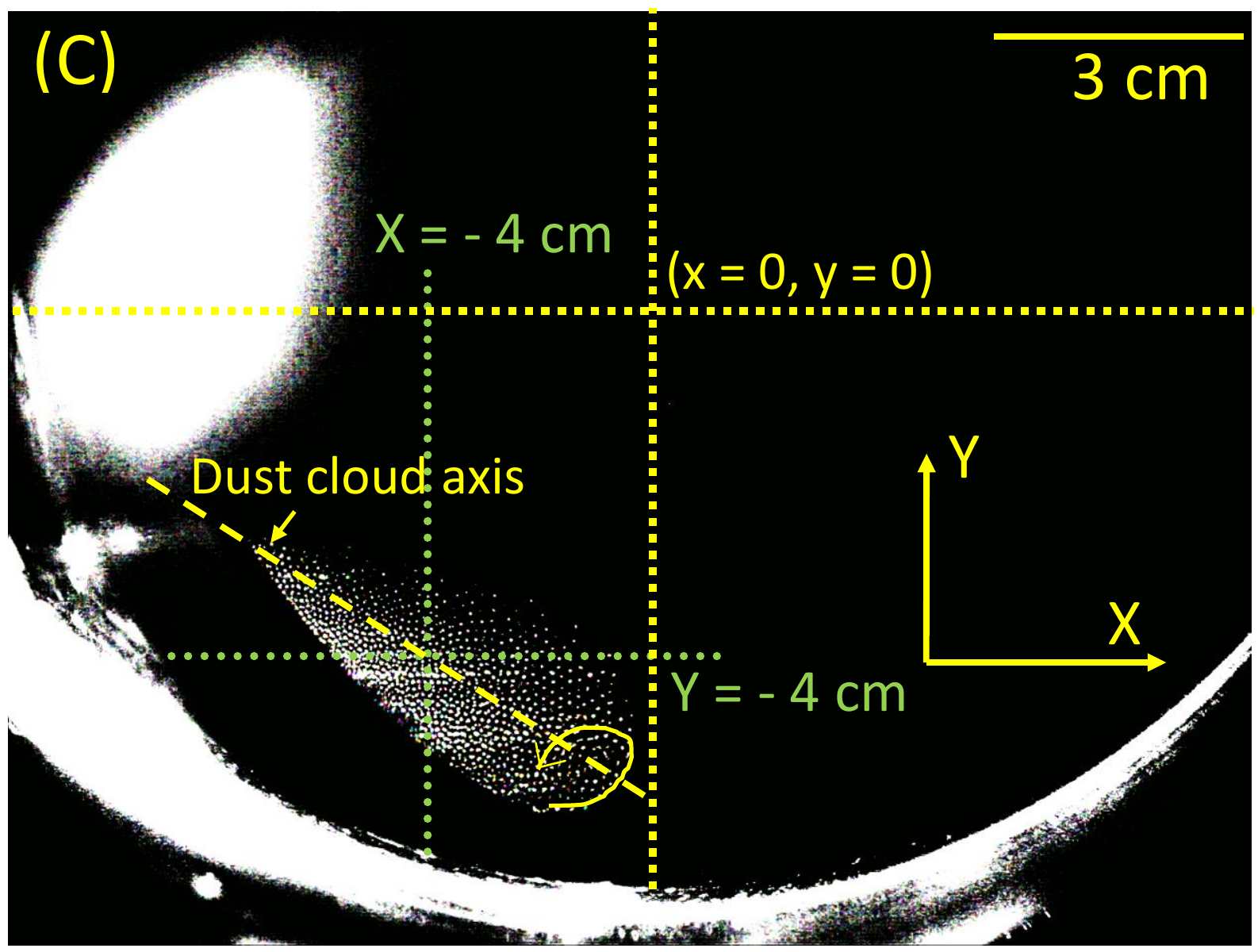}}}%
  \caption{\label{fig:fig1}(a)~Schematic of experimental configuration (top view). (b) A snapshot of the diffused plasma in X--Y plane at $\sim$12 $cm$ argon gas pressure p = 0.04 $mbar$ and RF power P = 6 W. White closed loop represents the boundary of the left axial port (or left view port) and Yellow dotted line represents the boundary of the diffused plasma. Confined dust particles at the bottom of the glow is indicated by white spots. (c) A full image of the confined dust cloud in X--Y plane near the center of source tube (Z $\sim$ 12 cm). Dust particles rotation is indicated by a yellow line with arrow. The circular arc (due to the reflected light) indicates the inner boundary of glass chamber. Position of dust cloud in this plane can be determined by using the reference of yellow dotted lines. Green dotted lines represent the typical measurements axis (along X and Y axis) for given Y and X values. Yellow dashed line represents the axis of the confined dust cloud in this plane.}
 \end{figure*}

The experiments are performed in a cylindrical linear device, which is described elsewhere \cite{mangilalrsi} in details. The schematic of the experimental assembly with an operating configuration is depicted in Fig.~\ref{fig:fig1}(a). In this experimental configuration Z = 0 $cm$ and Z = 60 $cm$ correspond to the left and right axial ports (as shown in Fig.~\ref{fig:fig1}(a)), respectively. X = 0 $cm$ and Y = 0 $cm$ indicate the points on the axis passes through the centre of the experimental tube. The centre of source tube is located at Z $\sim$ 12 $cm$, whereas the dust reservoir (a stainless steel disk of 6 cm diameter with a step like structure of 5 $mm$ width and 2 $mm$ height at its periphery) is located at Z $\sim$ 45 $cm$. A rotary pump, attached to buffer chamber, is used to evacuate the experimental chamber at $ \sim 10^{-3}$~mbar. Afterwards the argon gas is fed into the chamber till the pressure attains the values of $\sim$ 4--5 $mbar$. Then the chamber is pumped down again to the base pressure. This process is repeated three to four times to reduce the impurities from the vacuum chamber. Finally the operating pressure is set to 0.04 $mbar$ by adjusting the gas dosing valve and pumping speed.
%%%%%%%%%%%%%%%%%%%%
 %%%%%%%%%%%%%%%%%%%%%%%%%%%
A loop antenna (5 turns of copper wire) is wounded on the cylindrical source tube (Z $\sim$ 12 $cm$ long and 8 $cm$ diameter) as indicated in Fig.~\ref{fig:fig1}(a). Plasma is produced in the source tube using the 13.56 MHz RF generator, which later diffuses in the main experimental chamber. A snapshot of the  diffused plasma glow in X--Y plane for a given plasma parameters is shown in Fig.\ref{fig:fig1}(b). The boundary of glow region, as indicated in Fig.~\ref{fig:fig1}(b), changes with input RF power and gas pressure (not shown in the figure).\\
The diffused plasma in the main experimental chamber is characterized thoroughly by different electrostatic probes namely, the single \cite{probemerlino} and double \cite{doubleprobemalter} Langmuir probes and emissive probe \cite{emissivesheehan} in a working range of RF power (4--10 W) at argon pressure $p$ = 0.04 $mbar$. Plasma parameters traces along X and Y--axes for working discharge parameters are depicted in the Sec.\ref{sec:dust vortice}.
\paragraph*{•}
For injecting the kaolin dust particles ($\rho_d \sim$ 2.6 $gm/cm^3$ and $r_d \sim$ 0.5 to 5 $\mu$m) into the confining potential well, a secondary DC plasma source is used \cite{mangilalrsi}. As these particles come into the plasma volume, they start to flow in the ambipolar electric field of diffused plasma and found to confine in the potential well created by the diffused plasma \cite{mangilalrsi}. These confined dust particles are illuminated by the combination of a tunable red diode laser (632 nm wavelengths, 1--100 $mW$ power and 3 $mm$ beam diameter) and cylindrical lens (plano-convex). The dynamics of the dust particles are then captured by a IMPREX make CCD camera having frame rate of 16 fps and spatial resolution of 2352 $\times$ 1768 pixels. A standard zoom lens of variable focal length (from 18 $mm$ to 108 $mm$) is also used for the magnification purpose during the experiments. Maximum field of view at minimum zoom mode is 150 $\times$ 75 $mm^2$ and it reduces to 25 $\times$ 12.5 $mm^2$ at maximum zoom mode. There is a provision to image the dust cloud either in the X--Y or in the X--Z planes by changing the orientation of the cylindrical lens, laser and camera. A video image of the full view of confined dust cloud in the X--Y plane at Z $\sim$ 12 $cm$ is shown in Fig.~\ref{fig:fig1}(c). The series of images are then stored into a high-speed computer and later analyzed with the help of ImageJ and Matlab based available PIV software.\par
%%%%%%%%%%%%%%%%%%%%%%%%%%%%%%%%%%%%%%
\section{Experimental Observations on dust vortices} \label{sec:plasma} 
\begin{figure*}%[h]
\centering
  \includegraphics[scale=0.820]{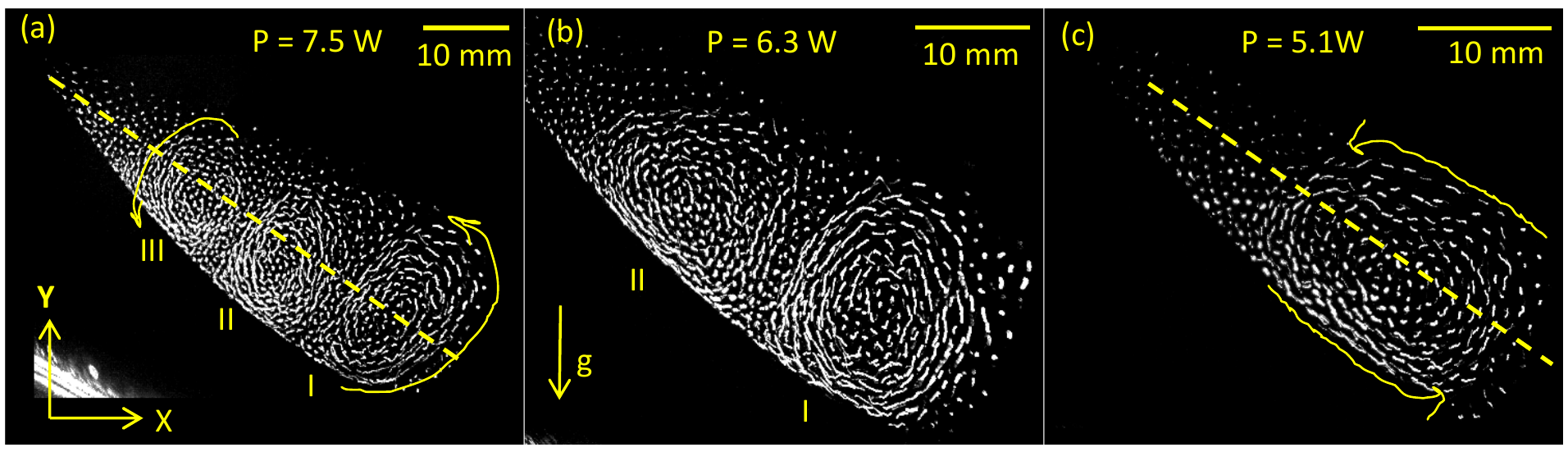}
%%%   %\vspace*{-0.13in}
\caption{\label{fig:fig2}Video images of dust cloud in the X--Y plane at Z = 12 cm. Images ((a)--(c)) are obtained with the superposition of five consecutive images at time interval of 66 ms. The dynamical structures in the extended dust cloud in this plane: Fig.\ref{fig:fig2}(a)--Fig.\ref{fig:fig2}(c) for input RF power (P) 7.5 W, 6.3 W, and 5.1 W respectively. Yellow lines with arrow indicates the direction of vortex motion of dust grains, arrow indicates the direction of gravity, and   dashed line corresponds to the axis of dust cloud. The vortex representation (I, II, and III) are made based on the number notation from the edge of the dust cloud. Kaolin particles are used to perform the dusty plasma experiments at argon pressure of 0.04 $mbar$.}
 \end{figure*}

As discussed in Section \ref{sec:exp_setup}, that dust particles are confined near the centre of source region (at Z $\sim$ 12 $cm$) and formed a 3D dusty plasma. Dynamics of the dust particles in X--Y plane at Z $\approx$ 12 $cm$ with different RF powers at fixed argon pressure $p$ = 0.04 $mbar$ is depicted in Fig.\ref{fig:fig2}. Five consecutive frames at time interval of 66 $ms$ are superimposed to get the information of the trajectories of the different particles. The contenious trajectories are seen in the Fig.\ref{fig:fig2}(a)--Fig.\ref{fig:fig2}(c) for the particles which follow a particular trajectory (or directed motion), whereas the randomly moving particles show only the dotted points. At RF power of P $>8$~W, the dust particles participate in the wave motion (similar to Vaulina \textit{et al.}\cite{selfexcitedmotion}), which is not shown in the figure. The dynamics of the particles when RF power is reduced at $P= 7.5$~ W is displayed in Fig.~\ref{fig:fig2}(a). At this discharge condition, almost all of the particles participate in rotational motion in the form of separated, co--rotating, anti--clockwise multiple (three) vortices. It is to be noted that the vortices are found to be stable until the discharge parameters are not changed. Small change in the ambient plasma parameters lead the distortion in the vortex structures. Further reduction of input power causes the reduction of dust cloud dimension (due to the fall down of the particles from the cloud edge) and as a result the inter--particle separation increases. At input power of P = 6.3 W, two co--rotating (anti--clockwise) dust vortices are observed in the dust cloud (see Fig.~\ref{fig:fig2}(b)). Dust cloud length and dust density decreases with further reduction of power to 5.1 W. At this discharge condition, the dust particles form a elongated single vortex structure (as shown in Fig.~\ref{fig:fig2}(c)) along the dust cloud axis. It is worth mentioning that dust particles near the mouth of the diffused plasma (nearby the plasma source) always exhibit the random motion.\\
%%%%%%%%%%%%%%%%%%%%%%%%%%%%%%%%%%%%%%%%%%%
 %%%%%%%%%%%%%%%%%%%%%%%%%%%%%%%%%%%%%
 %%%%%%%%%%%%%%%%%%%%%%%%%%%%%%%%%%%%%%%%%%%%%%%%%
 \begin{figure*}%[h]
\centering
  \includegraphics[scale=0.820]{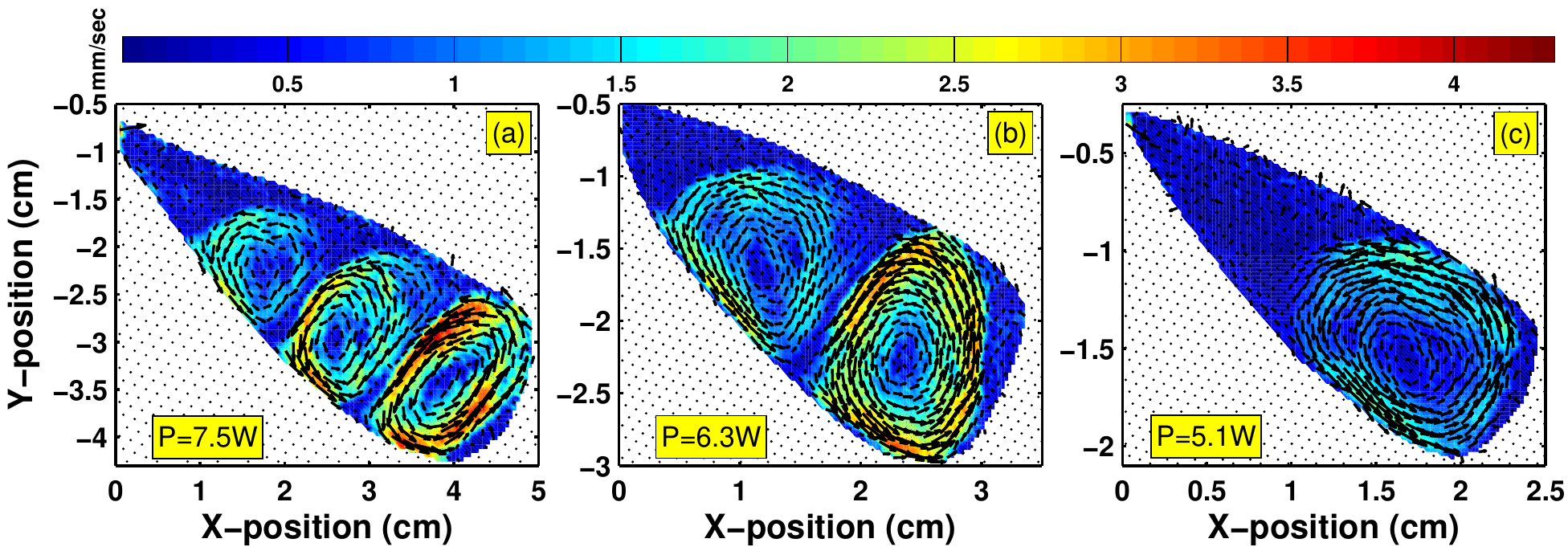}
%%%   %\vspace*{-0.13in}
 \caption{\label{fig:fig3}~ Images show the velocity distribution of dust particles in a vortex structure for different input RF powers (Fig.\ref{fig:fig2}). The images (Fig.\ref{fig:fig3} (a)--Fig.\ref{fig:fig3}(c)) are obtained after PIV analysis of the corresponding still images. Velocity vectors showing the direction of rotation of the dust particles in the X--Y plane at Z $\sim$ 12 cm. Color bar on the images show the value of the dust velocity in $mm/sec$. Three anti--clockwise co--rotating dust vortices are observed in the extended dust cloud at P = 7.5 W (Fig.\ref{fig:fig3}(a)). Dust cloud supports only two co--rotating vortices when input power is 6.3 W (Fig.\ref{fig:fig3}(b)). A single anti--clockwise dust vortex is observed at power of 5.1 W (Fig.\ref{fig:fig3}(c)) The argon gas pressure is kept fixed at 0.04 $mbar$}.
 \end{figure*}
 %%%%%%%%%%%%%%%%%%
 %%%%%%%%%%%%%%%%%%%%%%%%%%%%%
 \begin{figure}%[h]
\centering
\vspace*{-0.33in}
  \includegraphics[scale= 0.300]{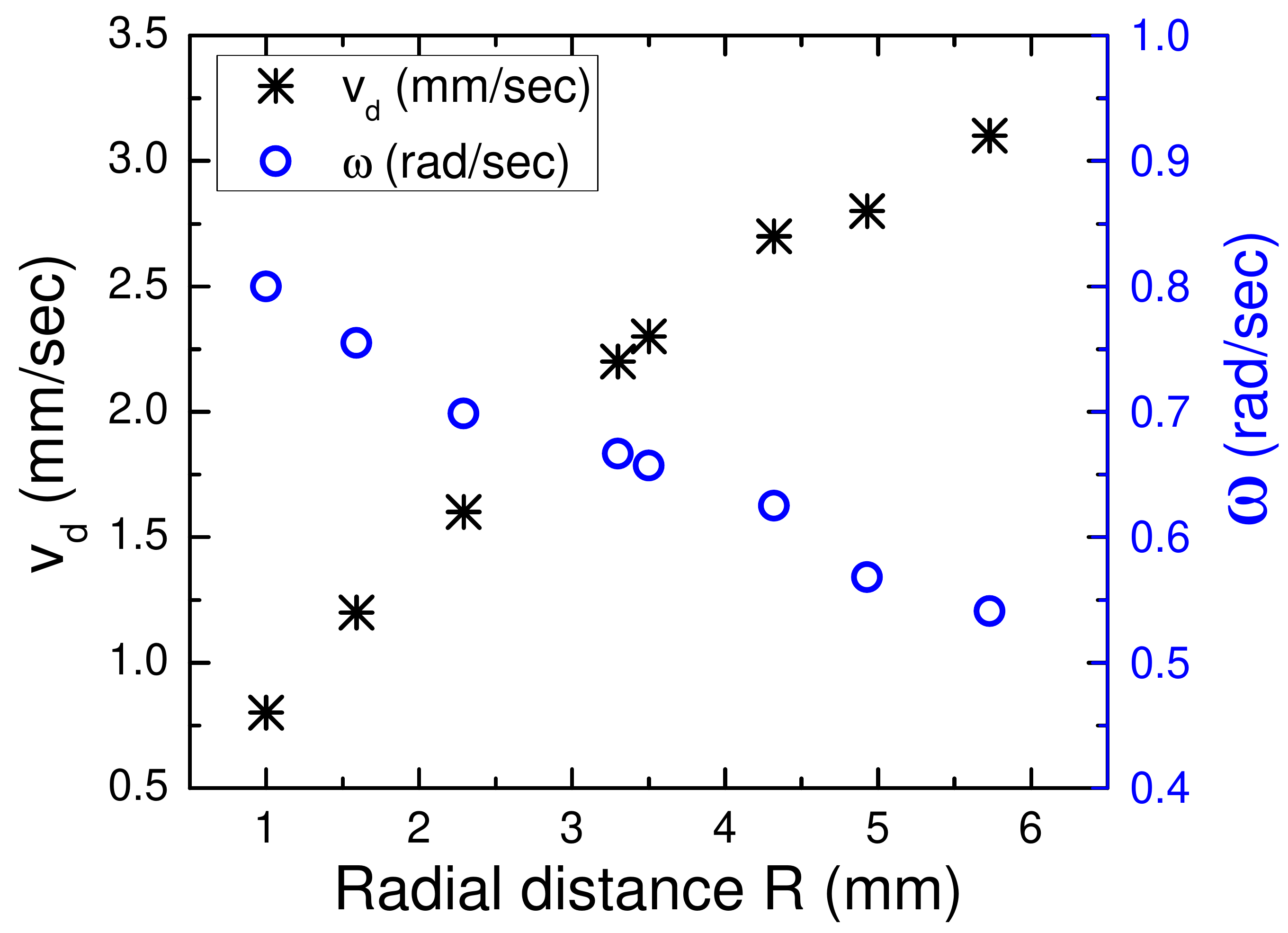}
\caption{\label{fig:fig4} Radial variation of rotation speed of particles and angular velocity for Vortex-I (Fig.~\ref{fig:fig2}(b)) at P = 6.3 W and p = 0.04 $mbar$.} 
 \end{figure}
 %%%%%%%%%%%%%%%%%%%%%%%%%%%%%%%
 \paragraph*{•}
MATLAB based open access software (Particles Image Velocimetry) called openPIV \cite{piv} is used to determine the direction and the magnitude of the particles velocities in each vortex structure. For constructing the vector field, an adaptive 2-pass algorithm (a 64$\times$64, 50\% overlap analysis, followed by a 32$\times$32, 50\% overlap analysis) is considered. The contour map of the average magnitude of the velocity, constructed after averaging the flow fields of 40 frames, is shown in Fig.~\ref{fig:fig3}. The direction of the field vectors (shown by arrows in the figure) represents the direction of particles motion in X--Y plane. The average velocity profile of the rotating particles shows that the velocity of particles is not uniform in a  particular vortex structure, as shown in Fig.~\ref{fig:fig3}(a). The particles rotate with minimum velocity at the centre of vortex, whereas it increases towards the boundary. The velocity profile of the three vortices (see Fig.~\ref{fig:fig3}(a)) at their boundary clearly shows that the particles in the biggest vortex (vortex--I) have maximum velocity and minimum in smallest vortex (vortex--III). It is worth mentioning that the interface (having anti--parallel flow) of two consecutive vortices are well separated. However, it is noticed that the magnitude of the particles velocity in vortex structure depends on the density of dust particles. The maximum velocity of the particles (at boundary) in the vortex structure decreases with the decrease of input RF powers as shown in Fig.~\ref{fig:fig3}. The maximum velocity (for vortex--I) is observed $\sim$ 4 $mm/sec$ and $\sim$ 2 $mm/sec$ at input power P = 7.5 W and P = 5.1 W, respectively. It is also to be noted that all the dust particles do not follow a closed path in the X--Y plane (some of the particles move to other plane) but the dust density remains nearly constant in the vortices. The radial distribution of the particle rotation speed and angular velocity ($\omega$) for vortex--I at P = 6.3 W is displayed in Fig.~\ref{fig:fig4}. It is observed that the particle rotational speed increases linearly towards the outer edge of the vortex. The rotational speed of particles is found to be higher with the large number of dust grains involving in formation of vortex structure. It is also seen that the angular velocity of rotating particles has a radial variation toward the outer boundary of the vortex. 
%%%%%%%%%%%%%%%%%%%%%%%%%%%%%%%
 \begin{figure*}%[h]
\centering
  \includegraphics[scale= 0.8200]{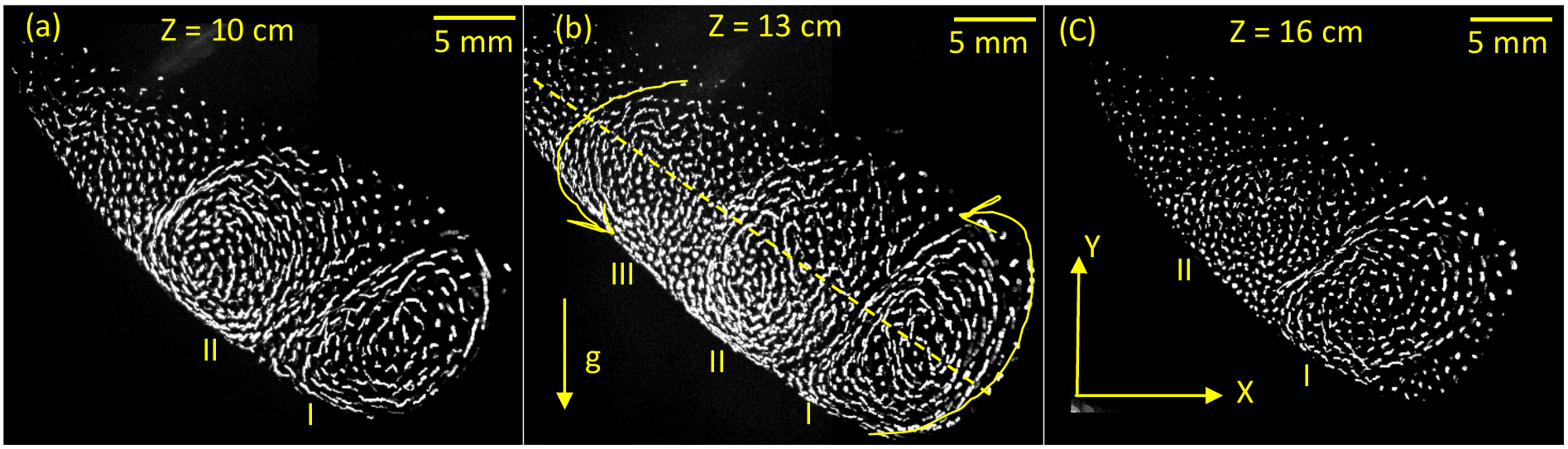}
%%%   %\vspace*{-0.13in}
\caption{\label{fig:fig5} Video images of dust cloud in different X--Y planes. All the images are obtained with the superposition of five consecutive images at time interval of 66 ms. Fig.\ref{fig:fig5}(a)-- Fig.\ref{fig:fig5}(c) corresponds to Z = 10 cm, Z = 13 cm and Z = 16 cm, respectively. Yellow lines with arrow indicates the direction of the rotating particles in the extended cloud. The RF power and gas pressure are 7.5 W and 0.04 $mbar$, respectively.}
 \end{figure*}
 %%%%%%%%%%%%%%%%%%%%%%%%%%%%%%%%%%%%
 \paragraph*{•}
 As the dust cloud is extended along the axis of the experimental chamber (along Z--axis), therefore it needs to investigate the dynamical structures in the different X--Y planes (i.e., at different Z location). Observation of the dust vortices in different planes  are shown in Fig.~\ref{fig:fig5}(a)--Fig.~\ref{fig:fig5}(c). It should be noted that the dust cloud has its maximum dimension near the center of the source region (Z $\sim$ 13 $cm$), which then decreases if one goes away from the source region. Number of vortex structures depends on the dimension of the dust cloud. For an example at P = 7.5 W and p = 0.04 $mbar$, three vortices are formed at Z $\sim$ 13 $cm$, whereas at other locations (Z $\sim$ 10 and 16 $cm$) only two vortices are observed. The velocity distribution of particles in the vortex structure is similar to that of described in Sec.~\ref{sec:dust vortice}.
\section{Origin of dust vortices} \label{sec:dust vortice} 
\begin{figure}
\centering
  \includegraphics[scale= 0.450]{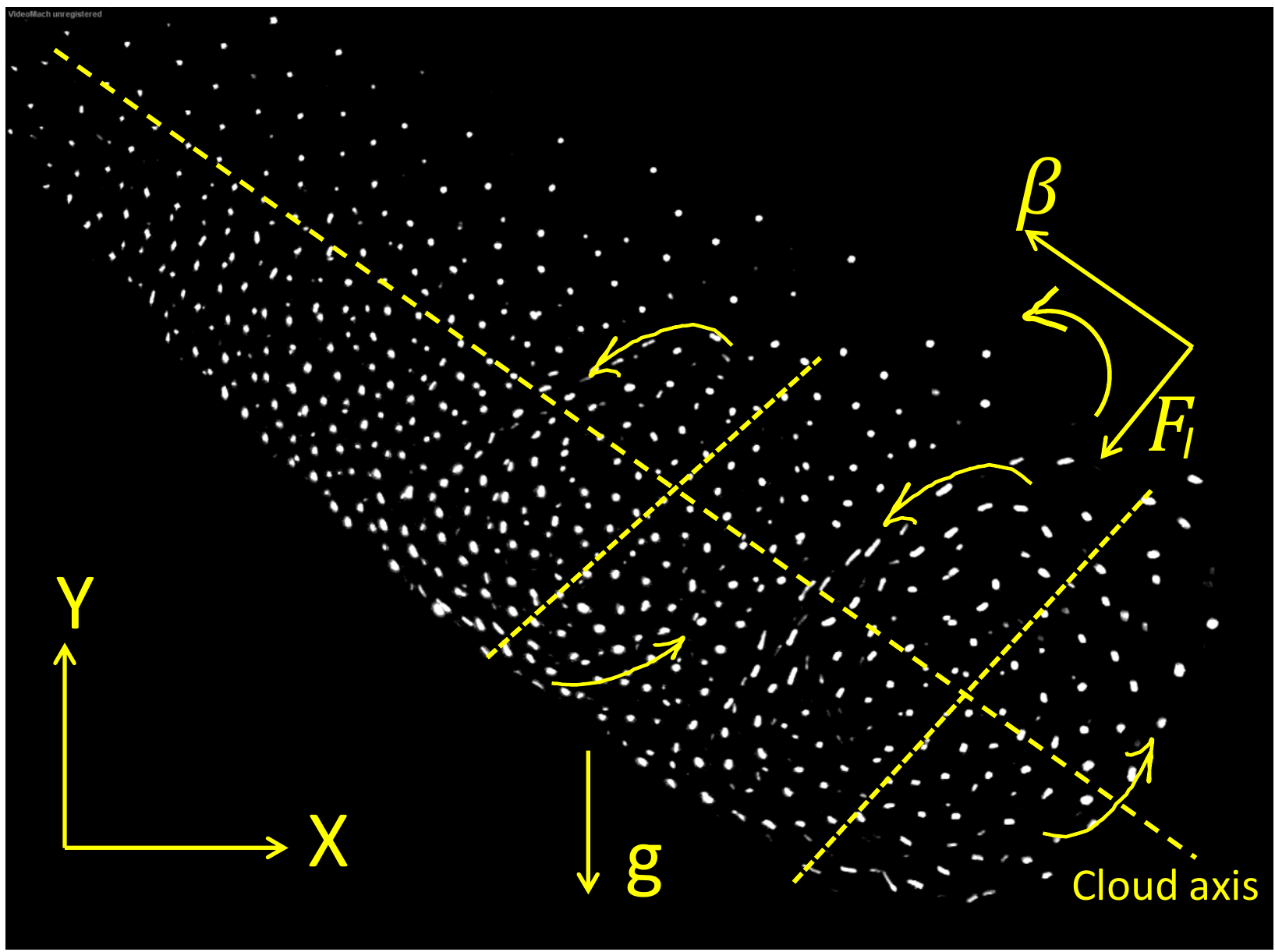}
%%%   %\vspace*{-0.13in}
\caption{\label{fig:fig6} Video image of dust cloud in the X--Y planes with direction of charge gradient ($\beta$) and ion drag force ($F_I$). The direction of rotation is displayed by a yellow line with arrow. Dust grains rotate in the direction of the gradient of dust charge.}
 \end{figure}

 For the formation of a steady-state equilibrium dust vortex, as described in Sec.~\ref{sec:plasma},  energy dissipation of the particles due to frequent dust--neutral collision and/or dust--dust interaction has to be balanced by the available free energy to drive the vortex motion. The spatial dependence of dust charge is one of the possible mechanisms to convert the potential energy into the kinetic energy of the dust particles \cite{vaulinajetp,vaulinaselfoscillation, selfexcitedmotion}, which results in to drive the vortex flow in a dusty plasma. The monotonic variation (gradient) of dust charge in a dusty plasma occasionally occurs due to inhomogeneity of the plasma parameters such as electrons (ions) density ($n_{e(i)}$) and/or electrons (ions) temperature ($T_{e(i)}$). In addition to that, non--dispersive nature of the dust particles sometimes also plays an important role to create a charge gradient in a dusty plasma. The present studies are carried out in a inductively coupled diffused plasma, where inhomogeneity in $T_e$ and $n_{e(i)}$ are expected, that causes the charge gradient along the length of dust cloud. Theoretical analysis and numerical simulations show such type of dynamical structures (vortices) in the presence of a dust charge gradient, $\vec{\beta} = \nabla Q_d = e\nabla Z_d$, orthogonal to a nonelectrostatic force $\vec{F}_{non}$ such as gravitational force ($\vec{F}_g$), ion drag force ($\vec{F}_I$), or thermophoretic force ($\vec{F}_{th}$) acting on the dust particles in the dust cloud \cite{vaulinajetp,vaulinaselfoscillation,selfexcitedmotion}. The role of non--electrostatic forces ($\vec{F}_{non}$) in the formation of dynamical structure in the dusty plasma is determined by their capacity to hold the particles in the region of non-zero electric field. 
Vaulina \textit{et al.} \cite{vaulinajetp, vaulinaselfoscillation,selfexcitedmotion} have carried an extensive study to explain the self oscillatory motion (acoustic vibrations, vortex etc.) in a dusty plasma with inhomogeneous plasma background. In such dusty plasma medium, they found that the curl of total force acting on the individual particle is non-zero due to a finite value of $\vec{\beta} \times \vec{E}$. In this case, the electric field does the positive work in compensating the dissipative energy losses. As a result an infinitely small perturbation, emerging in the dust cloud due to thermal and/or charge fluctuation, will grow in the system in the absence of restoring force and causes an instability in the dusty plasma, known as  dissipative instability \cite{selfexcitedmotion}. The evolution of this instability gives rise to a regular dynamic structures (vortices). The particle in dust cloud starts to move in the direction of $F_{non}$ where particle has its maximum charge value and form a vortex structure.
In the vortex motion, the vorticity ($\Omega = \nabla \times \vec{v}_d$) is non zero along a certain closed curve. The frequency ($\omega$) of the steady-state rotation of particles in a vortex structure is given by \cite{vaulinajetp,vaulinaselfoscillation,selfexcitedmotion},
 \begin{equation}
 \omega = \vert\frac{F_{non}}{M_d} \frac{\beta}{e Z_0 \nu_{fr}} \vert ,
 \end{equation}
 where $ Z_0 = Q_{d0}/e $ is charge on the dust particle at an equilibrium position in the rotating plane. In the present experimental configuration, dust cloud is confined in the X--Y plane as shown in Fig.~\ref{fig:fig1}(c). It is realized that the non--electrostatic force $\vec{F}_{non}$ required for the formation of the vortex motion of particle is induced by the directional motion of ions relative to the dust particles, i.e. $\vec{F}_{non} = \vec{F}_{I}$ (ion drag force) \cite{fortovvortex2}. Hence $\vec{F}_{non}$ can be replaced by $\vec{F}_I$ in the eq.(1) to obtain the angular frequency of the rotation. It should be noted that the force experienced by the particle due to gravity is found not to be orthogonal to charge gradient therefore its role on the vortex motion is not included in the calculations. However, its component along the ion drag force also contributes in the vortex motion. The schematic representation of the vortex motion in presence of charge gradient ($\beta$) and non--electrostatic force ($F_I$) in the X--Y plane is displayed in Fig.~\ref{fig:fig6}.
 According to Matsoukas and Russel's \citep{dustchargerusselapproxi.}, the charge on the dust grain ($Q_d$) can be expressed as:
\begin{equation}
\centering
Q_d = e Z_d \approx C \frac{4 \pi r_d k_B T_e}{e^2} ln \frac{n_i}{n_e}\left(\frac{m_e T_e}{m_i T_i}\right)^{\frac{1}{2}} ,
\end{equation}
 where $r_d$ is radius of the micro-particle, $k_B$ is Boltzmann's constant, $e$ is the electron charge, $n_e$ and $n_i$ are the electron and ion densities, $m_e$ and $m_i$ are their masses, and $T_e$ and $T_i$ are their temperatures. For a typical argon plasma, the constant comes out to be $C\approx$ 0.73 \citep{dustchargerusselapproxi.}. \par
Ion drag force $\vec{F}_I = F_I\hat{E}$, where $F_I$ is magnitude of the ion drag force which can be expressed as\cite{barnesdustforces}:
\begin{equation}
F_I = n_i v_s m_i v_i (\pi b^2_c + 4 \pi b^2_{\pi/2} \Lambda) ,
\end{equation}
 where $m_i$ is ion mass, $v_s$ is mean speed of ion, $v_i$ is ion velocity, $b_c$ is collection impact parameter \cite{barnesdustforces}, $b_{\pi/2}$ is the impact parameter whose asymptotic angle is $\pi/2$ and $\Lambda$ is the coulomb logarithm \cite{barnesdustforces}.   
The neutrals (either in rest or in motion) affect the motion of the dust particles in the plasma. In the present set of experiments, the direct gas flow inside the chamber is negligible \cite{mangilalrsi} thus neutrals are assumed in thermal equilibrium (or stationary). According to Epstein friction \cite{neutraldragepstein}, the neutral friction experienced by the dust particles can be given by 
 \begin{equation}
 \centering
 \vec{F}_n = -m_d \nu_{fr} \vec{v}_d,
 \end{equation}
 where $\nu_{fr}$ is the dust--neutral friction frequency and $v_d$ is dust particle velocity. The expression for $\nu_{fr}$ \cite{shukladustybook} is
 \begin{equation}
 \centering
 \nu_{fr} = \frac{8}{3} \sqrt{2 \pi} r^2_d \frac{m_n}{m_d} n_n v_{Tn} \left(1 +\frac{\pi}{8} \right) ,
\end{equation}  
 where $m_n$, $n_n$, and $v_{Tn}$  are the mass, number density, and thermal velocity of the neutral gas atoms, respectively.
 \par
%%%%%%%%%%%%%%%%%%%%%%%%%%%%%%%
 %%%%%%%%%%%%%%%%%%%%%%%%%%%%%%%%%%%%
 %%%%%%%%%%%%%%%%%%%%%%  
 \begin{figure*}
 \centering
\subfloat{{\includegraphics[scale=0.25]{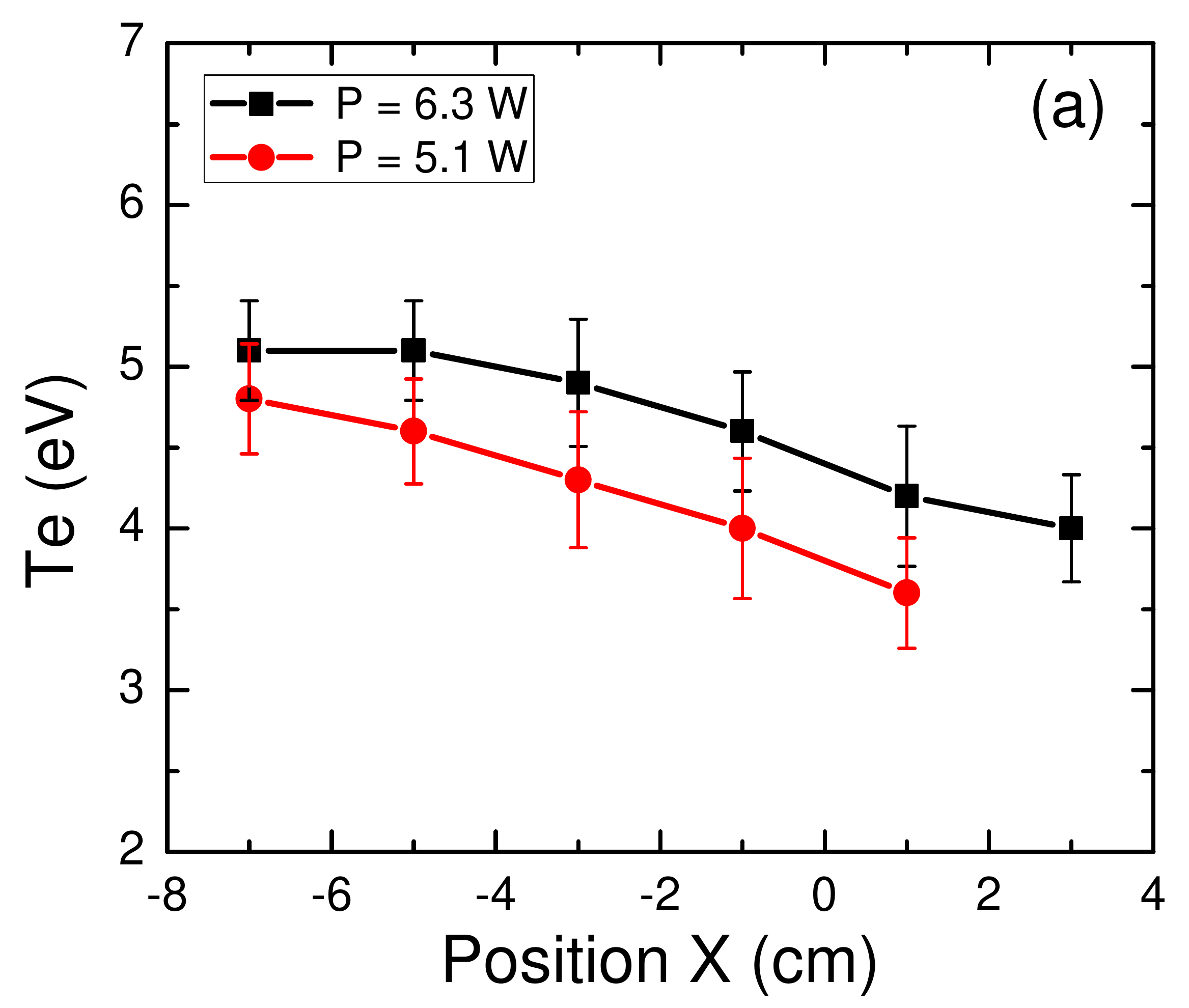}}}%
\hspace*{-0.4in}
 \qquad
%  %\vspace*{-0.in}
 \subfloat{{\includegraphics[scale=0.250]{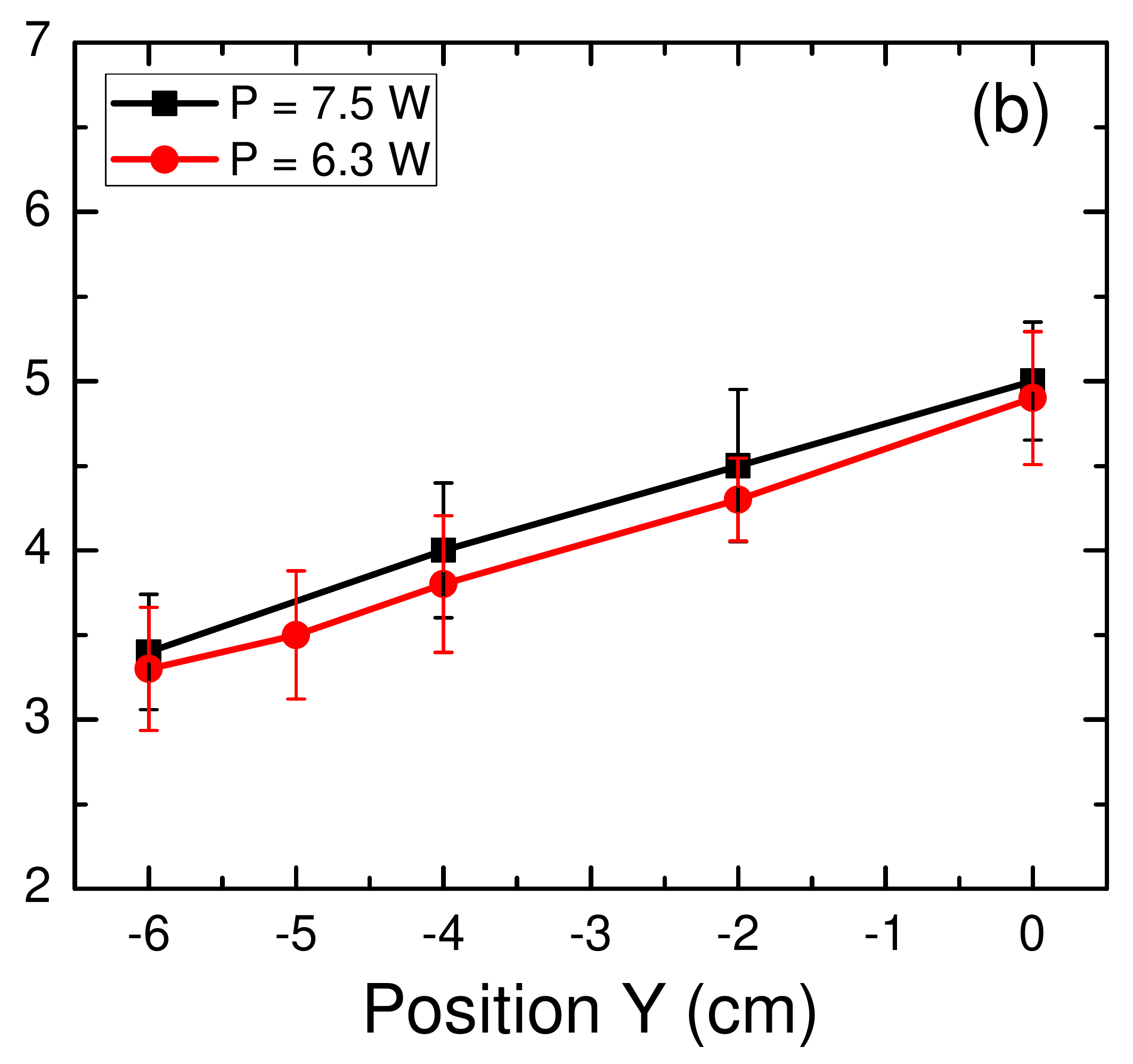}}}%p
 %\qquad
\hspace*{-0.05in}
\subfloat{{\includegraphics[scale=0.25]{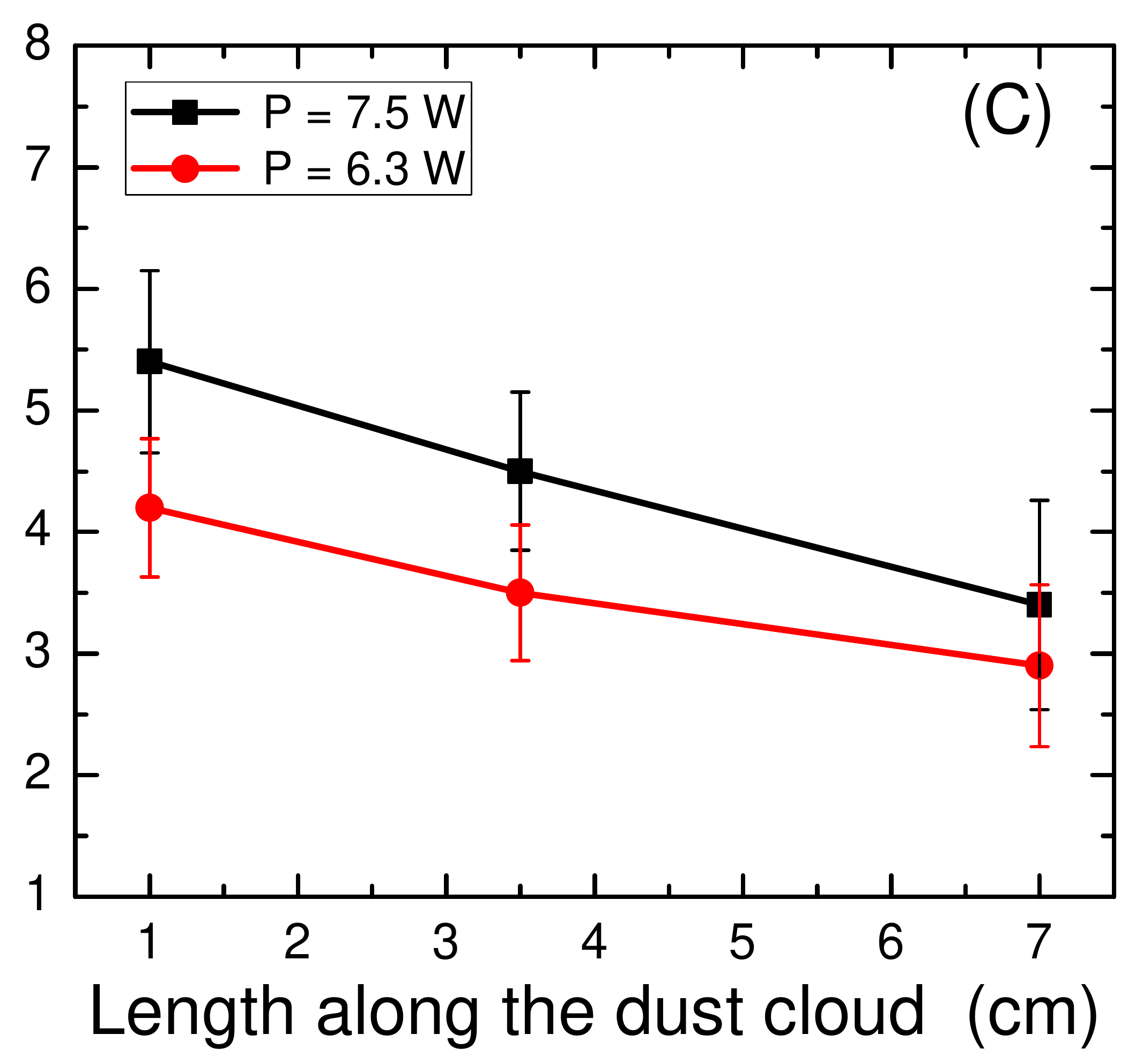}}}%
 \caption{\label{fig:fig7}(a) Electron temperature ($T_e$) variation along X--axis at Y = - 3 $cm$ and Z = 12 $cm$ for two RF powers P = 6.3 W and 5.1 W. (b) Electron temperature ($T_e$) variation along Y--axis at X = - 4 $cm$ and Z = 12 $cm$ for two RF powers P = 7.5 W and 6.3 W. (c) Electron temperature ($T_e$) variation along the axis of the confined dust cloud. The argon pressure is set at 0.04 $mbar$ during the experiments. All the measurements are taken in the normal plasma (without dust particles).} 
 \end{figure*}
 %%%%%%%%%%%%%%%%%%%%%%%%%%%%%%%%%%%%%%%%
 To estimate the angular velocity of dust rotation, it is necessary to estimate the dust charge gradient $\beta$ along the dust cloud axis and ion drag force $F_I$, which are assumed orthogonal to each other. It is described in the Sec.\ref{sec:plasma} that the dust cloud axis always lies in the X--Y plane, as shown in Fig.~\ref{fig:fig1}(c)). For calculating the dust charge gradient along the dust cloud axis, plasma parameters such as $n_e$ and $T_e$ are experimentally measured. In the present experimental configuration, it is very difficulty to trace the plasma parameters along the axis of the dust cloud. Therefore, the plasma parameters are scanned along the X--axis for given Y location and along the Y--axis for given X location at a particular Z position. These measurements are used to reconstruct the profiles of plasma density and electron temperature along the axis of the dust cloud. \par
\begin{figure*}
 \centering
\subfloat{{\includegraphics[scale=0.220]{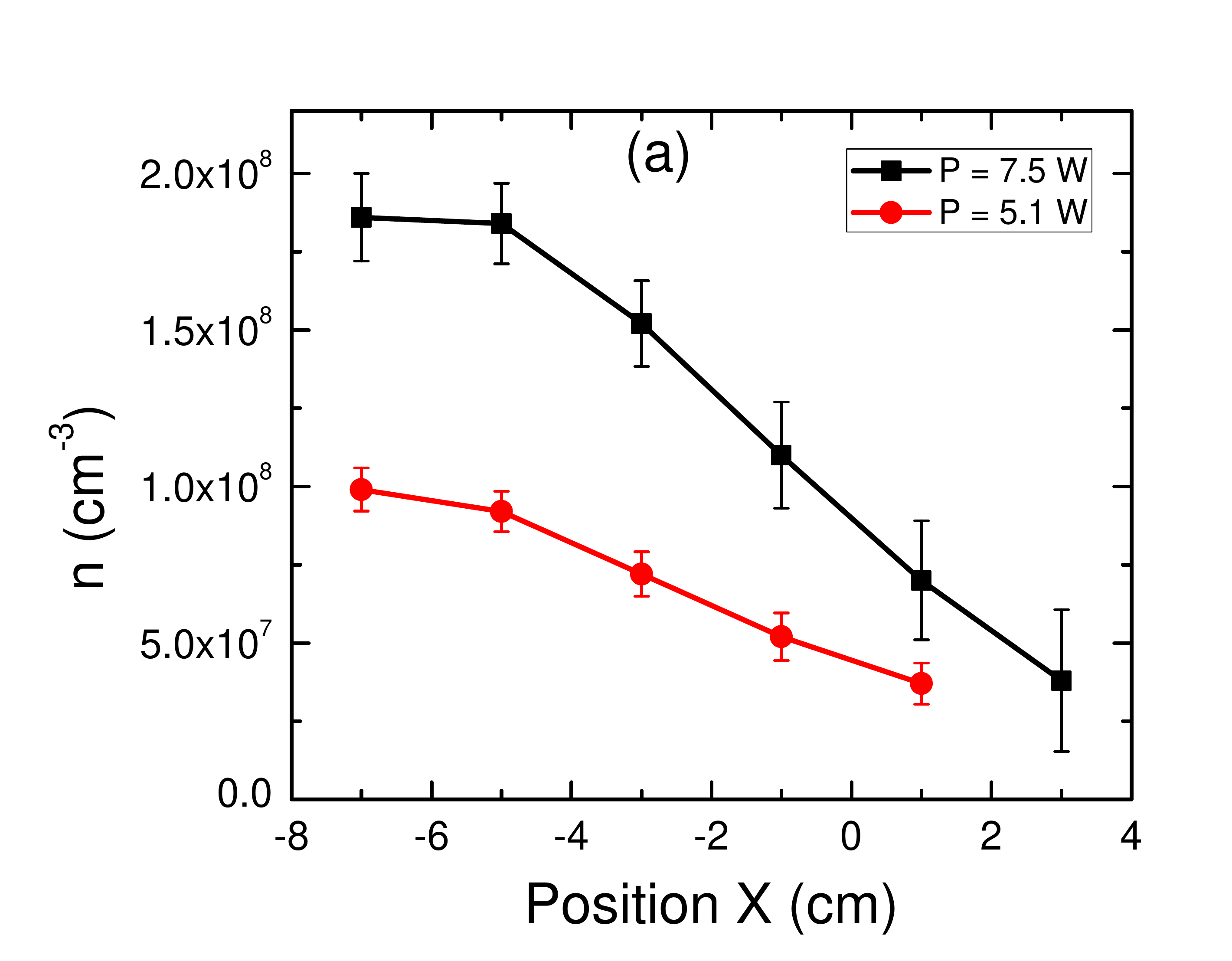}}}%
\hspace*{-0.3in}
 \qquad
%  %\vspace*{-0.in}
 \subfloat{{\includegraphics[scale=0.2200]{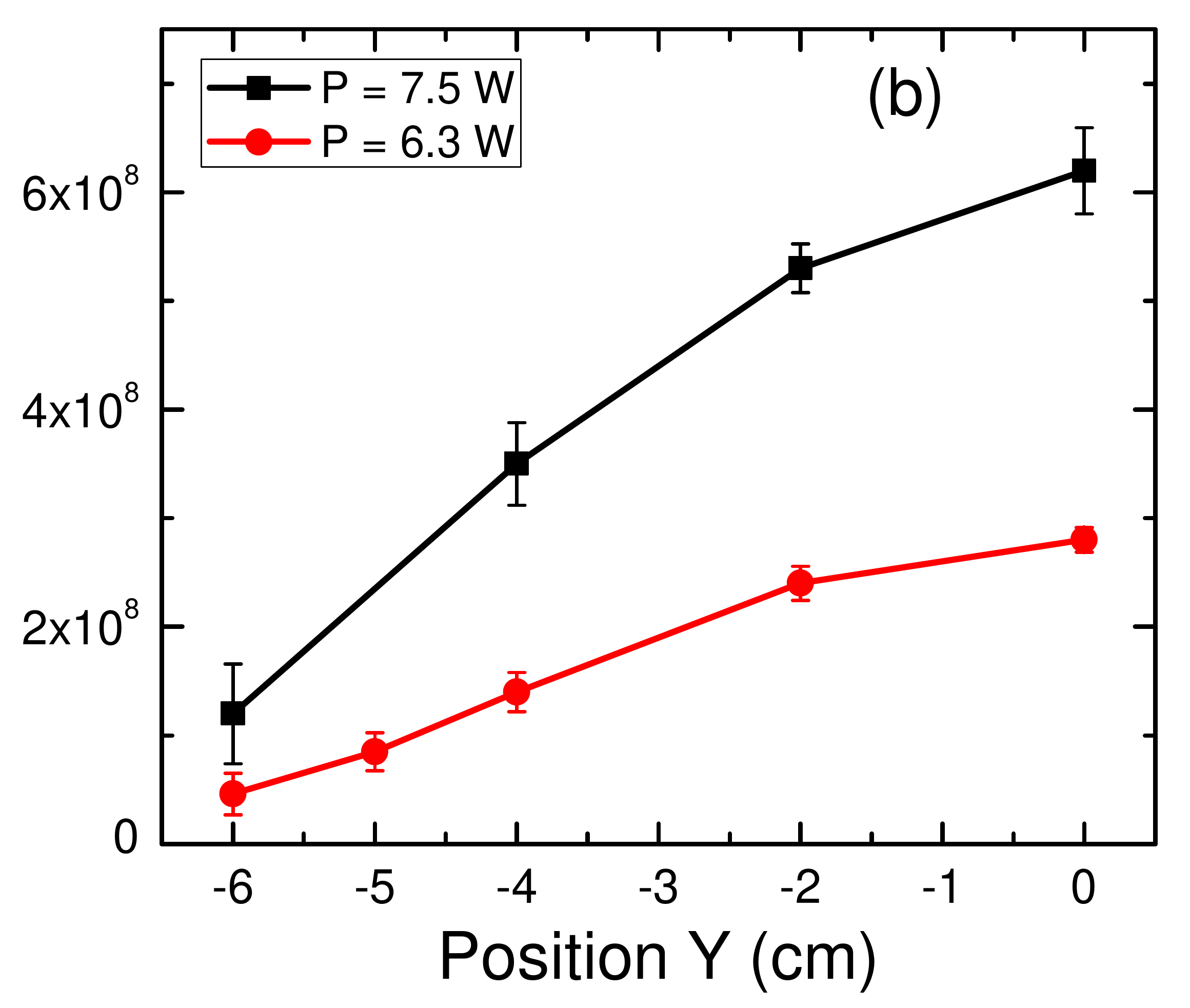}}}%p
 %\qquad
 \hspace*{-0.015in}
  %\vspace*{-05in}
\subfloat{{\includegraphics[scale=0.220]{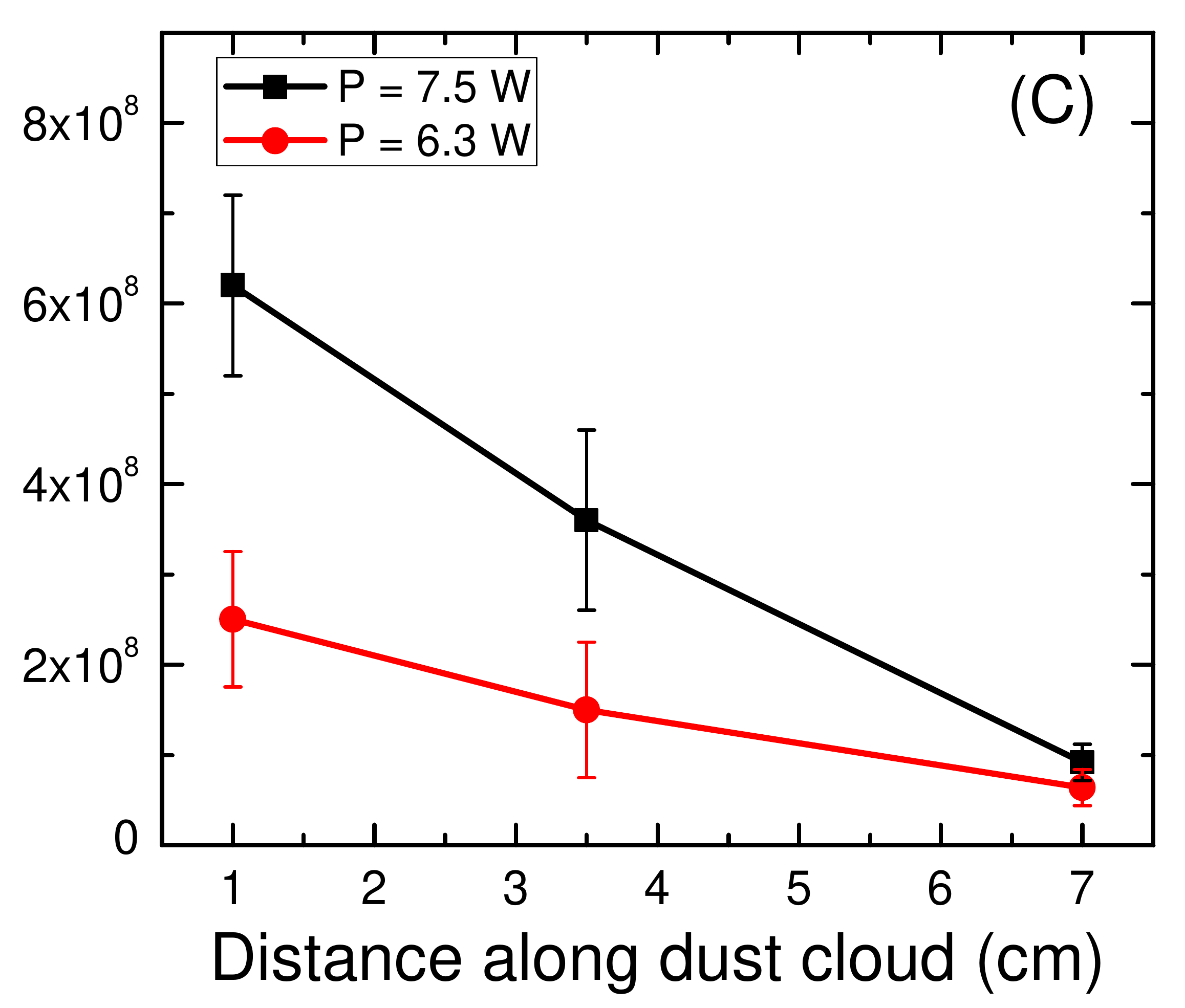}}}%
 \caption{\label{fig:fig8} (a)Plasma density ($n$) variation along X--axis at Y = - 3 $cm$ and Z = 12 $cm$ for two RF powers P = 6.3 W and 5.1 W. (b) plasma density variation along Y--axis at X = - 3 $cm$ and Z = 12 $cm$ for two RF powers P = 7.5 W and 6.3 W. (c) plasma density variation along the axis of the confined dust cloud for two powers P = 7.5 W and 6.3 W. The argon pressure is set at 0.04 $mbar$ during the experiments. All the measurements are taken in the normal plasma (without dust particles)}
 \end{figure*}

The variation of $T_e $ along the X--axis at Y = -3 $cm$ and Z = 12 $cm$ for different RF powers in absence of particles is shown in Fig.~\ref{fig:fig7}(a). It is seen in Fig.~\ref{fig:fig7}(a) that there is non--uniformity (or gradient) in $T_e$ along  X--axis for different RF powers. $T_e$ is observed to be high near plasma source (at X $\sim$ -7 $cm$) and decreases along the length of diffused plasma (from X = - 7 to X = 3 $cm$). Inhomogeneity in $T_e$ is also observed along the X--axis for different Y values (Y = 0 to - 5 $cm$). The variation of $T_e$ along the Y--axis at X = - 4 $cm$ and Z = 12 $cm$ is depicted in Fig.~\ref{fig:fig7}(b). It is clear from Fig.~\ref{fig:fig7}(b) that $T_e$ is also varied along Y--axis for different RF powers. Similar trend is also observed for different X--values (X = -7 $cm$ to X = 2 $cm$). It is also observed that $T_e$ increases with increase in the RF power (4 W to 10 W) at given X and Y locations. Using the different sets for $T_e$ profiles along X and Y--axes, Fig.~\ref{fig:fig7}(c) is constructed which shows the variation of $T_e$ along the axis of confined dust cloud. Fig.\ref{fig:fig7}(c) confirms that there exists a finite gradient in electron temperature along the dust cloud axis at a given RF power. Similar to $T_e$, plasma density $n$ is measured along the dust cloud axis. Typical plasma density variation along the X--axis at Y = -4 $cm$ and along Y--axis at X = -3 cm with different RF powers are shown in Fig.~\ref{fig:fig8}. It is found that the plasma density varies monotonically along the length of diffused plasma. It is higher near plasma source and decreases towards the chamber wall. Similar to previous case, the plasma density profile along the dust cloud axis as shown in Fig.~\ref{fig:fig8}(c) is extracted from the measured sets of density profiles along X axis at different Y values and along Y axis at different X values. Similar to $T_e$ there exist a density gradient along the dust cloud axis. For the estimation of dust charge and its gradient using Eq.~(2), an average radius of $\sim$ 2 $\mu$m is considered. According to this expression, there exists a charge gradient due to the presence of electron temperature gradient and plasma density gradient as shown in figure Fig.~\ref{fig:fig7}(c) and Fig.~\ref{fig:fig8}(c). However, the effect of density gradient on the dust charge gradient is negligible for a quasi-neutral plasma (see eq.~(2)). \par
 %%%%%%%%%%%%%%%%%%%%%%%%%%%%%%%%%%%%%%%%%%%%%%%%%%%%%%%%%%
\begin{figure}
 \centering
\subfloat{{\includegraphics[scale=0.3]{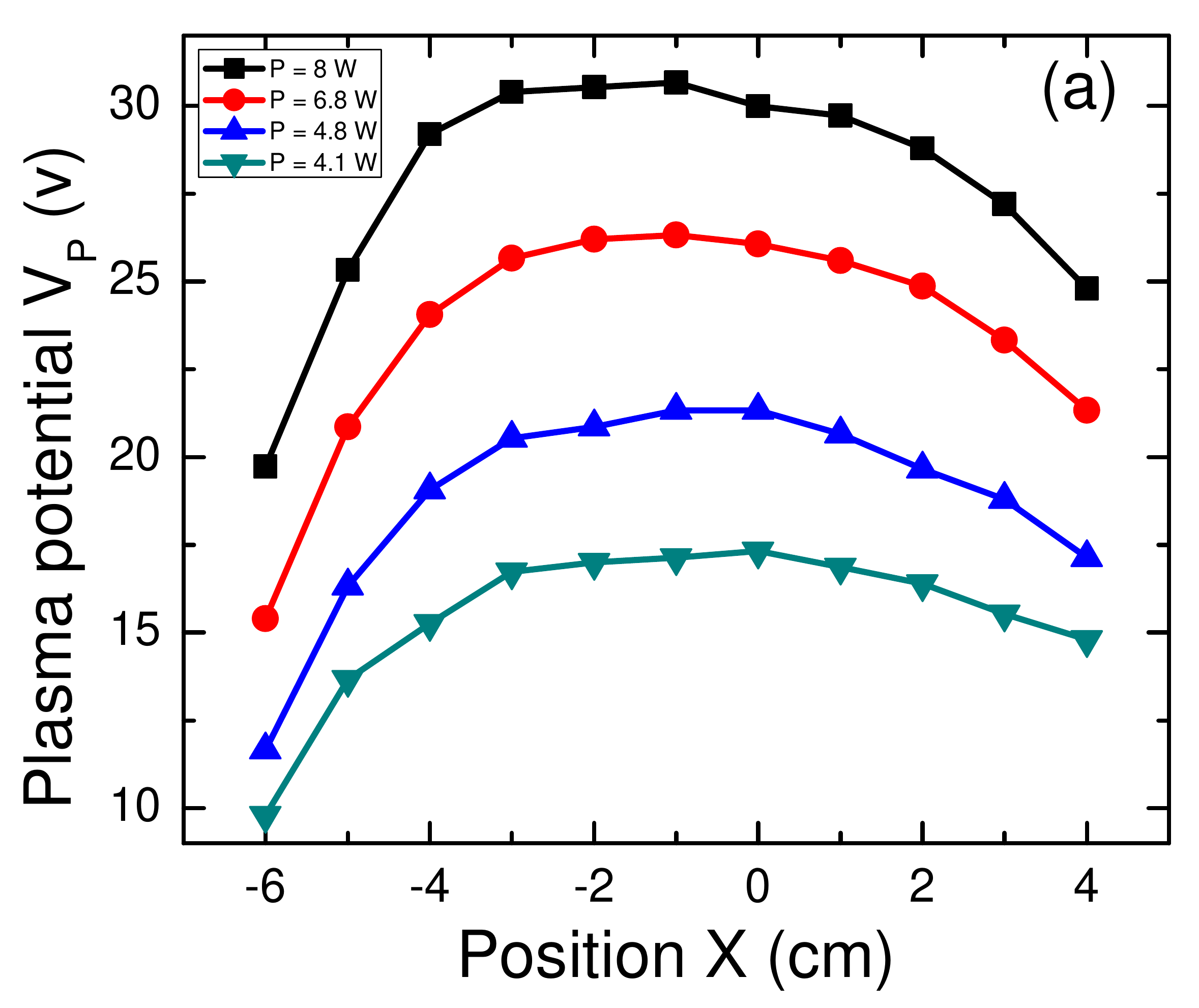}}}%
%\hspace*{-0.1in}
 \qquad
%  %\vspace*{-0.in}
 \subfloat{{\includegraphics[scale=0.3]{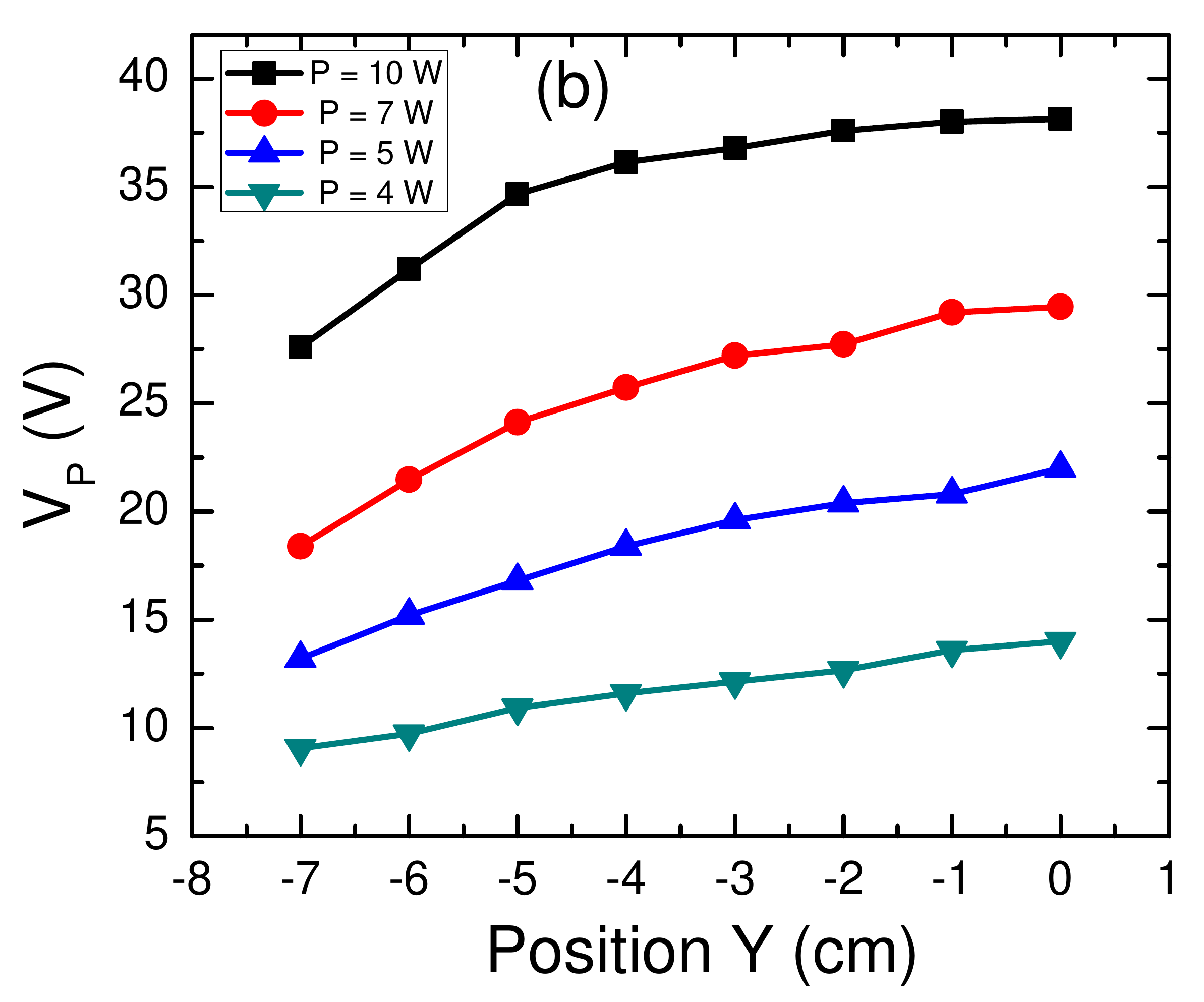}}}%p
 \caption{\label{fig:fig9} (a) Plasma potential profile along the X--axis at Y = - 4 $cm$ and Z = 12 $cm$ (b) along Y--axis at X = - 4 $cm$ and Z = 12 $cm$ for different RF powers. The plasma potential measurements are taken with emissive probe using floating point method. The argon pressure is fixed at 0.04 $mbar$ during the experiments. Errors in the measurements of plasma potential are within $\mp$ 2 V.}
 \end{figure}

 %%%%%%%%%%%%%%%%%%%%%%%%%%%%%%%%%%%%%%%%%%%%%%%%
For the estimation of electric fields along X and Y directions near the region where dust particles get levitated, an emissive probe is used to measure the plasma potential ($V_p$). The variation of plasma potential profiles are plotted in Fig.~\ref{fig:fig9} for different RF powers. Fig.~\ref{fig:fig9}(a) shows the potential profile along X--axis for Y = - 4 $cm$, whereas Fig.~\ref{fig:fig9}(b) shows the same along Y--axis for X = -4 $cm$ at centre of source tube (at Z $\sim$ 12 $cm$). It is to be noted that the plasma potential gradient near the glass wall (or diffused edge) is observed to be higher at higher RF power. The strong gradient gives higher E--field near the glass wall. The vertical E--field (along Y--axis) component holds the particles against gravity and the horizontal component (along X--axis) confines the particles as discussed in ref.\cite{mangilalrsi}. As shown in the Fig.~\ref{fig:fig9}(a), the X--component of E--field is negligible (flat $V_p$) inside the dust cloud (from X= -3 to 1 $cm$) and it has finite value at both the boundaries of the dust cloud. It is worth mentioning that the direction of E--field ($\hat{E}$) is perpendicular to the curved glass wall as shown in Fig.~\ref{fig:fig1}(c), which is orthogonal to the dust cloud axis (or along the direction of charge gradient).\par
 %%%%%%%%%%%%%%%%%%%%%%%%%%%%% 
 %%%%%%%%%%%%%%%%%%%%%%%%%%%%%%%
 The charge gradient along the axis of the dust cloud is defined as $\nabla Q_d = (Q_{d2} - Q_{d1})/(d_2-d_1)$,
 where $d_1$ and $d_2$ are the two spatial points on the dust cloud axis. For a quantitative analysis, only average sized particles ($\sim$ 2 $\mu$m) are considered based on the force balance conditions. As shown in Fig.~\ref{fig:fig4}, the observed value of angular frequency ($\omega_{exp}$) at P = 6.3 W and p=0.040~mbar is found to be in the range of $\sim$ 0.5--0.8 rad/sec. Theoretical estimated value of angular frequency ($\omega_{th}$) comes out to be $\sim$ 0.5 rad/sec for $\beta/e Z_0 \sim  $ 0.08 $cm^{-1}$, $M_d \sim$ 8 $\times 10^{-14}$ kg, $F_I \sim$ 7 $\times 10^{-14}$ N and $\nu_{fr} \sim$ 8 $sec^{-1}$. Similarly, $\omega_{exp}$ and $\omega_{th}$ are found to be $\sim$ 0.6--0.9 rad/sec and 0.4--1.1 rad/sec at P = 7.5 W, respectively. At lower power (P = 5.1 W), experimentally measured angular frequency $\omega_{exp} \sim$ 0.5--0.6 rad/sec is comparable with the estimated angular frequency $\omega_{th}  \sim$ 0.4--0.6 rad/sec. It can be concluded that the measured values of angular frequency and the theoretically predicted values (by Vaulina \textit{et al.}\cite{vaulinajetp}) are in good agreement for different RF powers. Moreover, the direction of rotation of the observed dust vortex is also consistent with the direction predicted in their theoretical model.  \par
The characteristic size $D_0$ of vortices can be obtained from the viscosity ($\eta_k $) of the dusty plasma medium \cite{vaulinasripta2004}, as $D_0 = \alpha \left(\eta_k / (\omega^* + \nu_{fr})^{1/2} \right)$, where $\omega^*$ is effective dusty plasma frequency and $\alpha$ takes into the difference between viscosity in quasi--stationary and dynamic vortex structure. The coefficient is estimated as $\alpha \approx$ 49 \cite{vaulinasripta2004}. The variation of kinetic viscosity ($\eta_k$) with a wide range of discharge parameters and coupling parameter ($\Gamma$) is discussed by Fortov \textit{et al.}\cite{fortovviscosity2}. For the present set of experiments, effective coupling constant ($\Gamma^*$) \cite{vaulinasripta2004} has the values between 10 to 100 for the particle of size ($r_d) \approx$ 2 $\mu$m, inter--particle distance ($d) \approx$ 700--900 $\mu$m, particle temperature ($T_d) \approx$ 0.2--0.4 eV and dust charge ($Q_d) \approx$ 1--5 $\times 10^{-15}$ C. In this parametric regime, the kinetic viscosity $\eta_k$ is considered to be $\sim$ 0.01 to 0.04 $cm^2 s^{-1}$ similar to the value report in refs\cite{fortovviscosity2,nosenkoviscosity2}. The characteristic size ($D_0$) of the vortices (shown in Fig~\ref{fig:fig2}(a)) for the parameters: $\eta_k$ = 0.02--0.03 $cm^2 s^{-1}$, $\nu_{fr}\sim$ 8 $s^{-1}$, dust Debye length ($\lambda_D) \approx$ 130 $\mu$m and $\omega^* \approx$ 40 $s^{-1}$ comes out to be $\sim$ 11--15 $mm$ which is in close match with the experimentally measured vortex diameter ($12-17$ mm). As the dimension of dust cloud in this discharge condition is $L\sim 55$~mm, hence the formation of multiple ($n=L/D_0 \sim 3$) vortex is possible to be accommodate in the dust cloud. In accordance with the above theoretical estimation, we also find three vortices to form in our experiments. As discussed in the Sec.~\ref{sec:dust vortice}, the length of dust cloud reduces with lowering the RF power by loosing the particles. The length of dust cloud reduces from $\sim$ 55 $mm$ to $\sim$ 25 $mm$ at P = 5.1 W. In addition to that, the average vortex size along the axis of the dust cloud increases to $\sim$ 16--19 $mm$ (see Fig.~\ref{fig:fig2}(c). For this discharge condition, the characteristic size ($D_0$) of the vortex is estimated as $\sim$ 11--16 $mm$ for $\eta_k \approx$ 0.02--0.04 $cm^2 s^{-1}$, $\omega^* \approx$ 30 $s^{-1}$ and $\Gamma^* \sim$ 60, which agrees well with the observed average size of the vortex. Therefore in this case, the number of vortex becomes $n=L/D_0 \sim 1$ as seen the experiment. The above preliminary understanding of multiple vortices is discussed in the light of characteristic scale length of dust vortex. However, there may be a possibility to exist an expanding Helical trajectories of dusts and the camera, having limited depth of focus due to the limited illuminated zone, sees a 2D image as multiple-vortices of a 3D expanding helical structure. The detailed nature and the reason for multiple vortices are still under investigation through further experiments and will be reported in future. 
\section{Summary and conclusion}  \label{sec:conclusion}
In this paper, an experimental observation of the formation of multiple co--rotating dust vortices over a wide range of discharge parameters is reported. Inductively coupled RF glow discharge is initiated in the background of argon gas in the source section, which diffuses in the main experimental chamber. A secondary DC glow discharge plasma is struck to introduce the dust particles into the plasma. These charged particles are drifted in the ambipolar electric field of the diffused plasma and start to confine in the electrostatic potential well, where the particles are trapped in the E--field which is result of the diffused plasma (ambipolar E--field) and glass wall charging (sheath E--field). At a particular discharge conditions, well separated, co--rotating, anticlockwise multiple vortices are found in the extended dusty plasma medium in the X--Y plane. At moderate RF power, two vortex structures are observed in the dust column. Only one vortex formed in the dusty plasma medium at lower RF power. The rotation speed is found to be non--uniform throughout the vortex structure and increases towards the boundary of vortex. The angular frequency of the rotation based on the model provided by Vaulina \textit{et al.} \cite{vaulinajetp,selfexcitedmotion} is found in close agreement with the experimentally observed values, which essentially shows that charge gradient in the dust column orthogonal to ion drag force is a possible mechanism to drive the vortex flow. The vortex structure has a characteristic size in the dusty medium which mainly depends on the dusty plasma properties (dust--dust interaction, dust--neutral interaction, dust density etc.). The quantitative description shows that multiple co--rotating vortices are probably formed in the dusty plasma with inhomogeneous plasma background when the vortex size is smaller than the dust cloud dimension. However, the detailed nature and the reason for multiple vortices are still under investigation and will be reported in the future publications.
\section{Acknowledgement} 
The authors grateful to Dr. M. Bandyopadhyay for his invaluable inputs to improve the manuscript. The author thanks Dr. D. Sharma and Dr. S. Ghosh for their valuable suggestions and discussions during the experiments. 
\bibliography{aipsamp}
\end{document}